\documentclass[reprint,superscriptaddress,
amsmath,amssymb,
pra,
]{revtex4-2}
\usepackage[utf8]{inputenc}

\usepackage{graphicx}
\usepackage{dcolumn}
\usepackage{bm}
\usepackage{hyperref}
\usepackage{ulem}
\usepackage{xcolor}
\usepackage{amsmath, amssymb}
\usepackage[T1]{fontenc}

\newcommand\Tstrut{\rule{0pt}{2.6ex}}   
\renewcommand{\vec}[1]{{\boldsymbol{#1}}} 

\usepackage{braket}
\usepackage{tikz}
\usepackage{pgfplots}
\usepgfplotslibrary{fillbetween}

\begin{document}
	

	
	\title{Dilute quantum liquid in a K-Rb Bose mixture}

	\author{V. Cikojevi\'c}
	\affiliation{University of Split, Faculty of Science, Ru\dj era Bo\v{s}kovi\'ca 33, HR-21000 Split, Croatia}
	\affiliation{Departament de F\'{\i}sica, Universitat Polit\`ecnica de Catalunya, Campus Nord B4-B5, E-08034 Barcelona, Spain}

	\author{E. Poli}
	 \affiliation{
     Institut f\"{u}r Experimentalphysik, Universit\"{a}t Innsbruck, Austria}
	\affiliation{Dipartimento di Fisica e Astronomia "Galileo Galilei" and CNISM, Universita di Padova, via Marzolo 8, 35122 Padova, Italy}
	
	\author{F. Ancilotto}
	\affiliation{Dipartimento di Fisica e Astronomia "Galileo Galilei" and CNISM, Universita di Padova, via Marzolo 8, 35122 Padova, Italy}

	\author{L.  Vranje\v{s}-Marki\'c}
	\affiliation{University of Split, Faculty of Science, Ru\dj era Bo\v{s}kovi\'ca 33, HR-21000 Split, Croatia}
	
	\author{J. Boronat} 
	\affiliation{Departament de F\'{\i}sica, Universitat Polit\`ecnica de Catalunya, Campus Nord B4-B5, E-08034 Barcelona, Spain}

	\date{\today}
	

	\begin{abstract} 
		A quantum liquid in a heterogeneous mixture of $^{41}$K and $^{87}$Rb 
atoms is studied using the diffusion Monte Carlo method and Density Functional 
Theory. The  perturbative Lee-Huang-Yang term for a heterogeneous 
mixture is verified and it is proved to be valid only near the gas-liquid 
transition. Based on the equations of state of the bulk mixture, calculated 
with  diffusion Monte Carlo, extensions to Lee-Huang-Yang corrected mean-field 
energy functionals (MF+LHY) are presented. Using Density Functional Theory, a 
systematic comparison between different functionals is performed, focusing on 
the critical atom number, surface tension, surface width, Tolman length, and 
compressibility. These results are given as a function of the inter-species 
interaction strength, within the stability domain of the liquid mixture. 
	\end{abstract}
	
	\maketitle

\section{Introduction}

A new quantum state of matter has been predicted \cite{petrov2015quantum}, and 
experimentally realized 
\cite{cabrera2018quantum,semeghini2018self,derrico2019observation} recently in 
ultracold atomic gases, where a subtle interplay between 
inter-species attractive interactions 
and quantum fluctuations may result in the formation 
of self-bound, ultradilute droplets.
Such liquid droplets are fundamentally different from those in classical 
or Helium fluids, where they arise instead from the interplay between
the short-range repulsive and long-range attractive
components of the interatomic potential \cite{barranco2006helium}.
The existence of self-bound ultra dilute quantum droplets,
made of atoms of a binary mixture of Bose-Einstein condensates, was predicted by 
Petrov \cite{petrov2015quantum} and
has been experimentally confirmed shortly later \cite{cabrera2018quantum,semeghini2018self}.
These systems, whose peculiar properties are shared by other systems
like dipolar Bose gases \cite{Pfau,Ferlaino,PhysRevLett.126.025301},
are characterized by ultralow densities, orders of magnitude
lower than that of the prototypical quantum liquid, i.e., liquid helium.
Noticeably, quantum droplets exist at temperatures that are several orders of
magnitude lower than the freezing points of classical liquids.

Mean-field analysis predicts that binary mixtures of 
Bose-Einstein condensates become unstable against 
collapse when the attractive inter-species 
interaction overcomes the repulsive contact potential 
between atoms \cite{pethick2008bose}. However, in the utradilute 
liquid phase the mean-field collapse is avoided 
if beyond-mean-field first-order perturbative corrections,
in the form of the Lee Huang Yang energy functional \cite{lee1957eigenvalues, 
larsen1963binary}, are included. This correction is repulsive in 
nature, and thus stabilizes the system. 

The formation of heteronuclear quantum droplets 
in an attractive bosonic mixture of $^{41}$K and $^{87}$Rb
has been observed recently \cite{derrico2019observation}. 
At variance with the largely studied mixture of two hyperfine states of 
$^{39}$K, longer lived self-bound states are observed in the K-Rb mixture,
both in free space and in optical waveguides. 
The K-Rb mixture has proven to be robust even when the two components are 
exposed to different confining potentials. Such long-lived, self-bound droplets 
remain localized on a timescale of 
several tens of milliseconds, more than a factor of 10 larger than in the 
$^{39}$K  mixture \cite{semeghini2018self}.

The increased lifetime of this new liquid mixture allows not only for a more 
detailed experimental observation and characterization of
isolated droplets, but will also permit the observation
of more complex scenarios arising from the interactions
between self-bound droplets. Moreover, the observed reduction of three-body 
losses allows for the  realization of larger droplets.
Recently, collisions between two droplets have been proposed as 
a useful experimental tool to investigate the dynamical 
properties of self-bound systems \cite{ferioli_collision}.
When two such droplets approach each other with a given relative velocity, 
they can either merge in a single droplet (coalescence) 
or separate into two or more droplets after the collision
(bouncing/fragmentation) \cite{ourdraftcollisions}. These different outcomes 
depend on whether or not the surface tension is 
large enough to counterbalance the kinetic energy of 
the colliding drops. 
A different phenomenology is expected
for the coalescence dynamics of two droplets colliding at very 
small velocities, which, in analogy to previous 
studies on helium clusters, could be a 
probe of their superfluid properties. One interesting outcome of
collision would be 
the formation of vortices or other topological structures during the 
merging process, as a consequence of their condensate nature.
These effects are known to arise 
during the merging of superfluid liquid helium nanodroplets \cite{Escart_n_2019}.

A fruitful comparison with experiments needs accurate and
reliable theoretical schemes, which have not been applied yet on K-Rb mixture. The use of
Density Functional Theory (DFT), in its time-dependent version, 
is known to allow a quite accurate description of the dynamics 
of inhomogeneous superfluid systems, even at the 
level of the Local Density approximation.
Finite-range effects permit to widen the applicability of numerical simulations 
by
going beyond the usual mean-field theory corrected with the LHY term (MF+LHY).
A most natural way to include such effects within the DFT scheme is 
to use density functionals built from results of first-principles
quantum Monte Carlo calculations for the homogeneous phase 
\cite{cikojevic2020finite, cikojevic2020qmcbased}. 
Importantly, first-principles finite-$N$ QMC calculations in the low-density 
regime show good agreement with the DFT approach \cite{parisi2020quantum} only 
in the vicinity of the gas-liquid transition while, at larger densities, 
modifications to MF+LHY approach are necessary \cite{petrov2015ultradilute, 
ota2020beyond, hu2020consistent, cikojevic2018ultradilute}.

This work is organized as follows. In Sec. \ref{sec:mflhy_eq_of_state}, we 
introduce  the MF+LHY functional for a K-Rb mixture and predict a range of scattering lengths which allow self-bound state. In Sec. 
\ref{sec:dmc_method}, we describe the diffusion Monte Carlo (DMC) methodology 
for obtaining the ground-state energies of the bulk mixture. We report the DMC 
energies obtained using the short-range and finite-range potentials in Sec. 
\ref{sec:short_range_potentials} and Sec. \ref{sec:realistic_potentials}, 
respectively. In Sec. \ref{sec:dft_results}, we apply the  
density-functional formalism to the characterization of the droplets, comparing 
the MF+LHY and QMC-based functionals for each quantity. Namely, we provide  
results for the surface tension, density profile, critical atom number, surface 
width, compressibility, and Tolman length for an experimentally relevant range 
of scattering parameters. Finally, a summary of results and conclusions is 
given in Sec. \ref{sec:summary}.

\section{\label{sec:mflhy_eq_of_state}MF + LHY equation of state}

We consider a homogeneous, heteronuclear Bose-Bose mixture with two components (with masses $m_1$ and $m_2$)
in a volume $V$ and a total number of bosons $N=N_1+N_2$.
By neglecting finite-range effects (which will be considered in  Sec. 
\ref{sec:dmc_method}), only the $s$-wave scattering
lengths are used to characterize the inter-particle
interactions. Within the LHY-extended mean-field framework (abbreviated MF+LHY hereafter), the energy of the 
system per unit volume is given by the functional \cite{petrov2015quantum}
	\begin{equation}
	\mathcal{E} = \mathcal{E}_{\rm MF} + \mathcal{E}_{\rm LHY} \ ,
	\end{equation}
where the mean-field (MF) and Lee-Huang-Yang (LHY) terms 
read \cite{pethick2008bose,Ancilotto_2018self,minardi2019effective}
	\begin{equation}
	\mathcal{E}_{\rm MF} = \dfrac{1}{2}g_{11}\rho_1^2 +  
\dfrac{1}{2}g_{22}\rho_2^2 + g_{12}\rho_1 \rho_2 \ ,
	\end{equation}    
	\begin{equation}\label{lhy_endens}
	\mathcal{E}_{\rm LHY} = \dfrac{8 m_1^{3 / 2} \left(g_{11} \rho_1  \right)^{5 
/ 2}}{15 \pi^2 \hbar^3}  \left[1 + \left(\dfrac{m_2}{m_1}\right)^{3/5} 
\dfrac{g_{22} \rho_2 }{g_{11} \rho_1} \right]^{5/2} \ ,
	\end{equation}
Here, $\rho_i$, ($i=1,2$) are the number densities of each component of the 
mixture, normalized such that $\int _V \rho_i\,{\rm d}{\bf r} =N_i$,
and $g_{ij} = 2\pi\hbar^2 a_{ij} / \mu_{ij}$ are the $ij$-interaction 
strengths, with $\mu_{ij}^{-1} = m_i^{-1} + m_{j}^{-1}$ the reduced mass. 

The mixture of the two species is stable against fluctuations in the 
concentration $N_1/N_2$ if \cite{petrov2015quantum}
	\begin{equation}
	\dfrac{\rho_2}{\rho_1} = \sqrt{\dfrac{g_{11}}{g_{22}}} \ .
	\end{equation}
		As pointed out in Refs. \cite{petrov2015quantum,Ancilotto_2018self, 
staudinger2018self}, it is safe to assume that this optimal composition is 
realized everywhere in the system. Thus, the energy functional 
$\mathcal{E}_{\rm MF+LHY}[\rho] = \mathcal{E}_{\rm MF}[\rho] + \mathcal{E}_{\rm LHY}[\rho]$ 
becomes effectively single-component, and can be written in terms of the total density 
$\rho = \rho_1 + \rho_2$. Under this assumption, the MF and LHY terms read
	\begin{eqnarray}
	\mathcal{E}_{\rm MF} & = &\frac{2 \pi \hbar^{2}  \left(2 a_{11} m_{2} + a_{12} \sqrt{\frac{a_{11} m_{2}}{a_{22} m_{1}}} \left(m_{1} + m_{2}\right)\right)}{m_{1} m_{2} \left(\sqrt{\frac{a_{11} m_{2}}{a_{22} m_{1}}} + 1\right)^{2}} \rho^{2} \hspace{0.2cm}  \\
	\mathcal{E}_{\mathrm{LHY}} &= &\frac{256 \sqrt{\pi} \hbar^{2} a_{11}^{5/2}}{15 m_1} \nonumber \\& \times & \left[ \frac{1+\left(\dfrac{m_2}{m_1}\right)^{{1}/{10}} \sqrt{a_{22} / a_{11}}}{1+\sqrt{\dfrac{m_2 a_{11}}{m_1 a_{22}}}}\right]^{{5}/{2}} \rho^{5/2}
	\end{eqnarray}

The MF+LHY energy per particle can be compactly written as
	\begin{equation}
	\label{eq:eos_mflhy}
	\dfrac{E/N}{|E_0^{\rm MF+LHY}| /N} = -3 \dfrac{\rho}{\rho_0^{\rm MF+LHY}} + 
2\left(\dfrac{\rho}{\rho_0^{\rm MF+LHY}}\right)^{3/2}  \ ,
	\end{equation}
$\rho_0^{\rm MF+LHY}$ being the equilibrium density within the MF+LHY theory,
i.e., the density that minimizes the functional $\mathcal{E}_{\rm 
MF+LHY}[\rho]$. Explicitly,
	\begin{eqnarray}
	\label{eq:rho_0_mflhy}
	\rho_0^{\rm MF+LHY} =&  \displaystyle \frac{25 \pi m_{2}^{3} \left(2 a_{11} m_{2} + a_{12} \sqrt{\frac{a_{11} m_{2}}{a_{22} m_{1}}} \left(m_{1} + m_{2}\right)\right)^{2} }{4096 \left(a_{11} m_{2} + a_{22} m_{1} \left(\frac{m_{2}}{m_{1}}\right)^{\frac{3}{5}} \sqrt{\frac{a_{11} m_{2}}{a_{22} m_{1}}}\right)^{5}}   \nonumber \\ &\times \left(\sqrt{\frac{a_{11} m_{2}}{a_{22} m_{1}}} + 1\right).
	\end{eqnarray}
The energy per particle at equilibrium is 
	$E_0^{\rm MF+LHY} / N = \mathcal{E}_{\rm MF + LHY}[\rho_0^{\rm MF+LHY}] / \rho_0^{\rm MF+LHY}$.
The fact that the MF+LHY functional can be written in a universal form (Eq. 
\ref{eq:eos_mflhy}) proves that all the results from 
Ref. \cite{petrov2015quantum} can be applied here with a proper change of units.
	
In what follows, we define the hyperfine state $\ket{F=1, m_F=1 }$ of $^{41}$K 
as 
component 1, and the hyperfine state $\ket{F=1, m_F=1 }$ of $^{87}$Rb 
as component 2. The scattering parameters describing the intra-species repulsion 
are fixed and their values are equal to $a_{11} = 65 a_0$ \cite{derrico2007feshbach} 
and $a_{22} = 100.4 a_0$ \cite{marte2002feshbach}. With those two parameters, 
the MF+LHY theory predicts a self-bound state (hereafter called "liquid") 
for $a_{12} < a_{12}^c$, with $a_{12}^c$ given by
	\begin{equation}
	a_{12}^c = \dfrac{-2 \sqrt{a_{22} / a_{11}}}{\sqrt{m_2 / m_1} \left(1 + m_1 / m_2\right)} \approx -75.4 a_0.
	\end{equation}
In the experiment, accessible values of $a_{12}$ are in the 
range between $a_{12} = -80a_0$ and $a_{12} = - 95a_0$ \cite{derrico2019observation}.

\section{\label{sec:dmc_method} Diffusion Monte Carlo}
	
We use the diffusion Monte Carlo (DMC) method to determine the energy per 
particle in the homogeneous phase. This method was previously applied in related 
problems regarding the study of a Bose-Bose liquid by some of the authors 
\cite{cikojevic2019universality,cikojevic2020finite}. DMC is nowadays a 
well-know method that is able to solve exactly the imaginary-time Schr\"odinger 
equation of the many-particle system, within some statistical noise. The starting 
point of  DMC is the decomposition of the imaginary-time evolution operator. 
In this paper, we use a propagator that is accurate up to second order in 
the timestep \cite{sarsa2002quadratic}, following  
the implementation outlined in Ref. \cite{boronat2002microscopic}. 
	
	To reduce the variance in the estimation of energy, we use standard 
importance 
sampling through a trial wavefunction, written as a Jastrow product over pairs 
\cite{jastrow1955many}
	\begin{equation}
	\Psi(\vec{\mathrm{R}}) = \prod_{j>i=0}^{N_1} f^{(11)}(r_{ij})  
\prod_{j>i=N_1}^{N_1+N_2} f^{(22)}(r_{ij}) \prod_{i,j}^{N_1, N_2} 
f^{(12)}(r_{ij})\ ,
	\end{equation}
	where the two-particle correlation functions $f^{(\alpha,\beta)}(r)$
($\alpha,\beta = 1, 2$) are chosen as
	\begin{equation}
	\label{eq:trial_wf}
	f^{(\alpha,\beta)}(r)=
	\begin{cases}
	f_{\rm 2b}(r) &  r < \tilde{R} \\
	B(1 - \frac{a_{\alpha,\beta}}{r}) & r < R_v , \\
	C\exp(-\frac{D}{r} + \frac{E}{r^2})       ,& R_v <r < L/2 \\
	1      ,& r > L/2 \ . \\
	\end{cases}
	\end{equation}
	The two-particle correlation function at short distances $f_{\rm 2b}(r)$ 
is the solution to the two-body problem for a given interaction potential. Throughout the paper we use short-ranged potentials, such that the potential is zero at distances greater than $\tilde{R}$. The function $f_{\rm 2b}$ is connected to the asymptotic form $1-a_{\alpha,\beta}/r$, where $a_{\alpha,\beta}$ is a corresponding scattering length.  At $r=R_v$, this is connected to the long-range phononic form $C\exp(-\frac{D}{r} + \frac{E}{r^2})$ \cite{reatto1967phonons}. Finally, we impose that the function  is constant (one)  at the boundary of the simulation box 
($r=L/2$). This trial wavefunction has only one variational parameter, 
namely $R_v$, which we optimize variationally finding that, in all cases,  
$R_v = 0.45 L$. Our DMC results are unbiased  for 
time-steps $\Delta \tau \lesssim 0.5 m_{41} a_{11}^2 / \hbar^2$ and walker 
number $n_{\rm w} \approx 200$. Simulations are performed in a cubic box of 
size $L=(N/\rho)^{1/3}$, with periodic boundary conditions applied to particle 
coordinates. For each density, a set of several calculations of increasing 
number of particles, namely $N=130$, 160, 200, 250, and 500 is performed in 
order to study finite-size effects. The final energies, corresponding to the 
thermodynamic limit,  are obtained  by an extrapolation to 
$N\rightarrow \infty$ assuming a correction  which decreases as 
$N^{-1}$.

	\section{\label{sec:short_range_potentials}Short-range potentials}
	
\begin{figure}[]
		\centering
		\includegraphics[width=\linewidth]{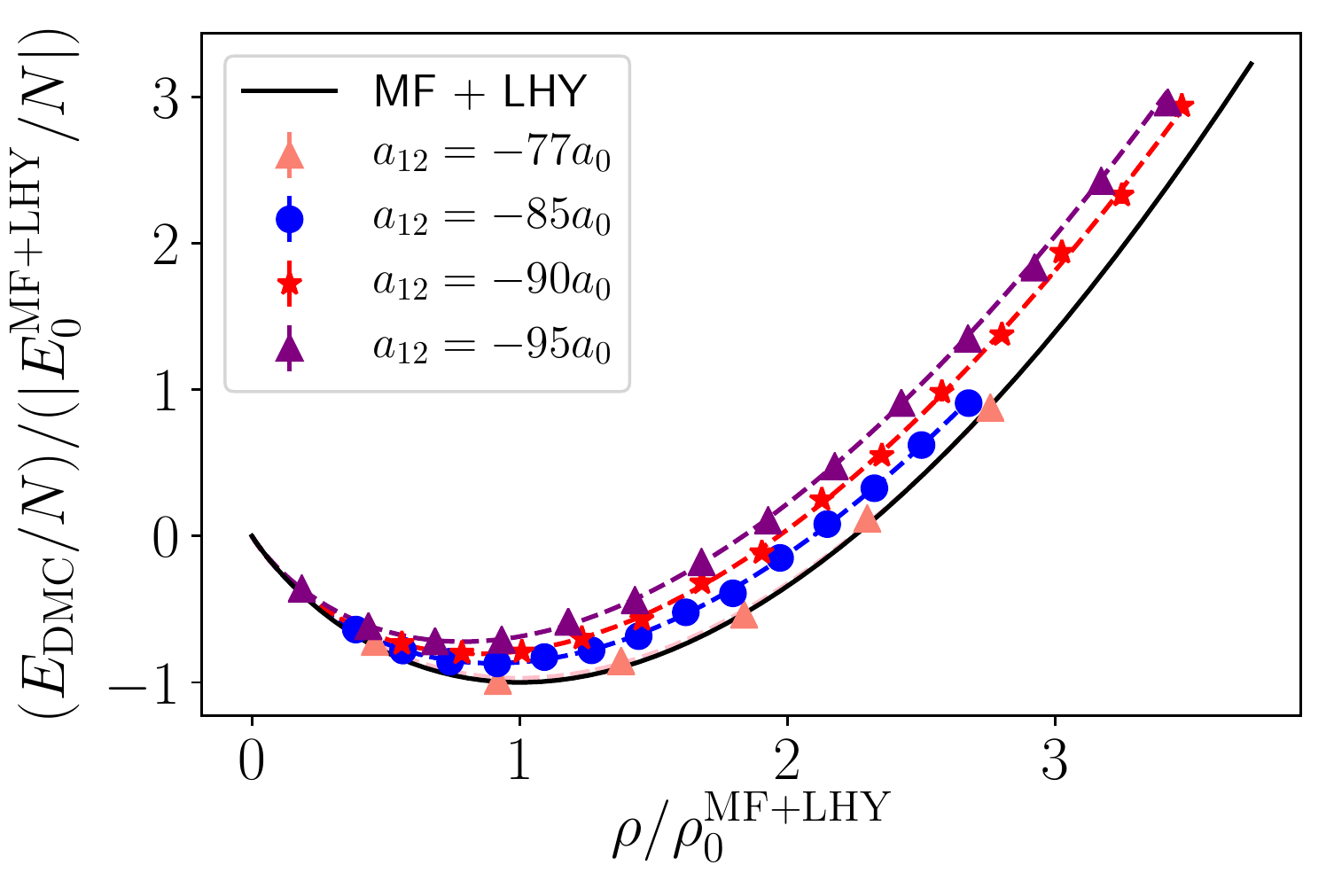}
		\caption{DMC energy per particle as a function of density, obtained with the POT-SR models (Eqs. \ref{eq:hard-core_potential} and \ref{eq:square-well_potential}), in a mixture having the optimal composition $\rho_2 / \rho_1 = \sqrt{g_{11} / g_{22}}$. Energy per particle and the total density are normalized with respect to the equilibrium values given in Eq. (\ref{eq:rho_0_mflhy}).}
		\label{fig:universaleos2onlyshortrange}
	\end{figure}

	In a first approach to the problem, we use in DMC interatomic 
potentials that have a very short range, which we call POT-SR in the following. 
Since we cannot use a contact interaction in DMC, we model the short-range 
interaction by a set of potentials with a range $r_p$ satisfying $\rho r_p^3 
\ll 1$, where $\rho$ is the typical number density. Under this criterium, we 
choose a hard-core potential for repulsive interactions (between equal 
species), with a diameter corresponding to the $s$-wave scattering length 
\cite{pethick2008bose}
	\begin{equation}
	V_{ii}(r)=
	\begin{cases}
	\infty  &  r < a_{ii},\\
	0, &  \mathrm{otherwise} \\
	\end{cases}, \hspace{0.5cm} i=1,2.
	\label{eq:hard-core_potential}
	\end{equation} 
	The attraction between different species is modeled by a short-range 
square-well potential, 
	\begin{equation}
	V_{12}(r)=
	\begin{cases}
	-V_{\rm sr}, & r <R_{\rm sr},\\
	0, & \mathrm{otherwise} \\
	\end{cases},
	\label{eq:square-well_potential}
	\end{equation} 
	where we choose $R_{\rm sr} = a_{11}$. The particular choice of potentials 
given in Eqs. (\ref{eq:hard-core_potential}) and 
(\ref{eq:square-well_potential}) resembles a zero-range case, since the 
probability of finding two particles within the diameter $R_{\rm sr}$ 
is $\rho_0 R_{\rm sr}^{3} \approx 2 \cdot 10^{-4}$, evaluated at the 
equilibrium density for the densest liquid analyzed, 
corresponding to $a_{12} = -95 a_0$. By properly setting $V_{\rm sr}$, we obtain 
a target scattering length $a_{12}$, which for a square-well potential is given 
by \cite{stoof2009ultracold}
	\begin{equation}
	a_{12} = R_{\rm sr} \left\{ 1 + \dfrac{\tan \left(K_0 R_{\rm sr}\right)}{K_0 
R_{\rm sr}}  \right\} \ ,
	\end{equation}
where $K_0 = \sqrt{2 V_{\rm sr} (m_1 + m_2) / (\hbar^2 m_1 m_2)}$. DMC results 
of the energy per particle, obtained with POT-SR potentials (Eqs. 
\ref{eq:hard-core_potential} and \ref{eq:square-well_potential}), are shown in 
Fig. \ref{fig:universaleos2onlyshortrange}, for $a_{12}=-77a_0$, $-85a_0$, 
$-90a_0$, and $-95a_0$, which include experimentally accessible values. They are compared with the predictions of MF +LHY 
theory. The DMC energy per particle is well fitted to the form
\begin{equation}
	\label{eq:dmc_en_fit}
	\dfrac{E}{N} = \alpha \rho + \beta \rho^\gamma \ ,
\end{equation}
where $\alpha$, $\beta$, and $\gamma$ are the fitting parameters. The 
equilibrium density and the coefficients of the fit to the DMC energy per 
particle are reported in Table \ref{table:fit_params_POT-SR}. As we can see in 
Fig. \ref{fig:universaleos2onlyshortrange},  the DMC results of the energy 
per particle reproduce well the MF +LHY theory for $a_{12}=-77a_0$, which is 
close to the critical value, $a_{12}^c=-75.4 a_0$. 
However, the DMC results for different $a_{12}$ do not fit a single line defined 
by Eq. (\ref{eq:eos_mflhy}), implying that the universality of the MF +LHY 
theory is broken, with the deviation growing as $a_{12}$ becomes larger. 
Interestingly, we observe repulsive beyond-LHY contributions when the 
fluid enters a more correlated regime. Repulsive beyond-LHY contributions to 
the energy have already been observed in symmetric Bose-Bose fluids 
\cite{cikojevic2019universality} and in the  liquid  $^{39}$K 
mixture \cite{cikojevic2020finite} for the potentials with a small effective 
range. This effect could be due to bosonic 
pairing between atoms of  different species \cite{hu2020consistent, 
ota2020beyond}.

	\begin{table}[]        
		\caption{Coefficients of fit $E/N = \alpha \rho a_{11}^3 + \beta (\rho a_{11}^3)^{\gamma}$ of the DMC energies per particle, obtained with short-range set of potentials POT-SR (Eqs. (\ref{eq:hard-core_potential}) and (\ref{eq:square-well_potential}), reported in Fig. (\ref{fig:universaleos2onlyshortrange}). Parameters $\alpha$ and $\beta$ are given in units $\hbar^2 / (m_1 a_{11}^2)=10^{-3}{\rm Kelvin}$, and $\gamma$ is adimensional. $\rho_0$ stands for the equilibrium density of a QMC functional. Scattering lengths in the repulsive channel are $a_{11} = 65 a_0$ and $a_{22} = 100.4 a_0$.}
		\label{table:fit_params_POT-SR}
		\centering 
		\begin{tabular}{ c | c | c | c | c | c}
			\rule{0pt}{3ex}  $~a_{12}~$ &    Method & $\alpha$ & $\beta$ & $\gamma$ & $\rho_0 a_{11}^3 $\\ \hline          
			\rule{0pt}{3ex}   $-77 a_0$ & QMC    & $-0.056$ & $35.709$ & $1.500$ & $1.08 \times 10^{-6}$ \\         
			\rule{0pt}{3ex}   & MF+LHY & $-0.057$ & $36.437$ & $1.5$ & $1.09 \times 10^{-6}$ \\ \hline
			\rule{0pt}{3ex}  $-85 a_0$ & QMC    & $-0.385$ & $15.981$ & $1.395$ & $3.48 \times 10^{-5}$ \\
			\rule{0pt}{3ex}  & MF+LHY & $-0.340$ & $36.437$ & $1.5$ & $3.86  \times 10^{-5}$ \\ \hline
			\rule{0pt}{3ex}   $-90 a_0$ & QMC    & $-0.592$ & $16.344$ & $1.384$ & $7.53  \times 10^{-5}$ \\
			\rule{0pt}{3ex}   & MF+LHY & $-0.516$ & $36.437$ & $1.5$ & $8.92  \times 10^{-4}$ \\ \hline
			\rule{0pt}{3ex}   $-95 a_0$ & QMC    & $-0.772$ & $16.768$ & $1.379$ & $1.26  \times 10^{-4}$ \\
			\rule{0pt}{3ex}   & MF+LHY & $-0.693$ & $36.437$ & $1.5$ & $1.61 \times  10^{-4}$ 
		\end{tabular}
	\end{table}

	\section{\label{sec:realistic_potentials}Finite-range potentials}
	
	The effective range in the $^{41}$K-$^{87}$Rb mixture is not known, but can 
be estimated from the combined knowledge of the $C_6$ coefficient 
of the leading term in the long-range tail of the Van Der Waals interaction potential
and the 
scattering length obtained from measurements of the Feshbach resonances 
\cite{chin2010feshbach, gribakin1993calculation, flambaum1999analytical}. 
Knowing the van der Waals coefficient $C_6$, one can estimate the 
effective range using a semiclassical 
approximation \cite{flambaum1999analytical},
	\begin{equation}
	r_{\rm eff} = \dfrac{\Gamma(1/4)^2}{6 \pi } a_m \left(1 - 2\dfrac{a_m}{a} + 
2 \left(\dfrac{a_m}{a}\right)^2\right) \ ,
\label{rangevdw}
\end{equation}
	with $a_m = {4\pi R_{\rm vdw}}/{\Gamma(1/4)^2} $ the mean scattering 
length and $R_{\rm vdw}= \left({2\mu C_6}/{\hbar^2}\right)^{1/4}/ 2$ the van 
der Waals length, where $\mu$ is the reduced mass. The $C_6$ coefficients for 
$^{41}$K and $^{87}$Rb are given, along with the effective ranges derived from 
(\ref{rangevdw}), in Table \ref{table:scattering_params}.
	
	Since only two scattering parameters, namely the scattering length $a$ and
the effective range $r_{\rm eff}$, cannot uniquely define the interaction 
potential, we resort to model potentials satisfying the two scattering criteria. 
To investigate the role of the "shape" of the interaction dictated by the 
higher-order scattering parameters, we performed two independent sets of 
calculations, each having different models of the interaction potential. We call 
the two models POT-I and POT-II, and both can be written as follows
	\begin{equation}
	V(r)=
	\begin{cases}
	V_0 & 0 <r <R_0,\\
	-V_1, & R_0 <r <R_1, \\
	0, & \mathrm{otherwise}. \\
	\end{cases} \hspace{0.5cm} i=1,2.
	\label{eq:realistic-potentails}
	\end{equation} 
	This particular form of interaction is convenient because the analytic 
expressions for both the $s$-wave scattering length and effective range are 
analytically known \cite{jensen2006bcs}. The specific values of the interaction 
parameters in all three channels for POT-I and POT-II potentials are summarized 
in Table \ref{table:realistic-pots_params}.
	
\begin{figure}[]
		\centering
		\includegraphics[width=\linewidth]{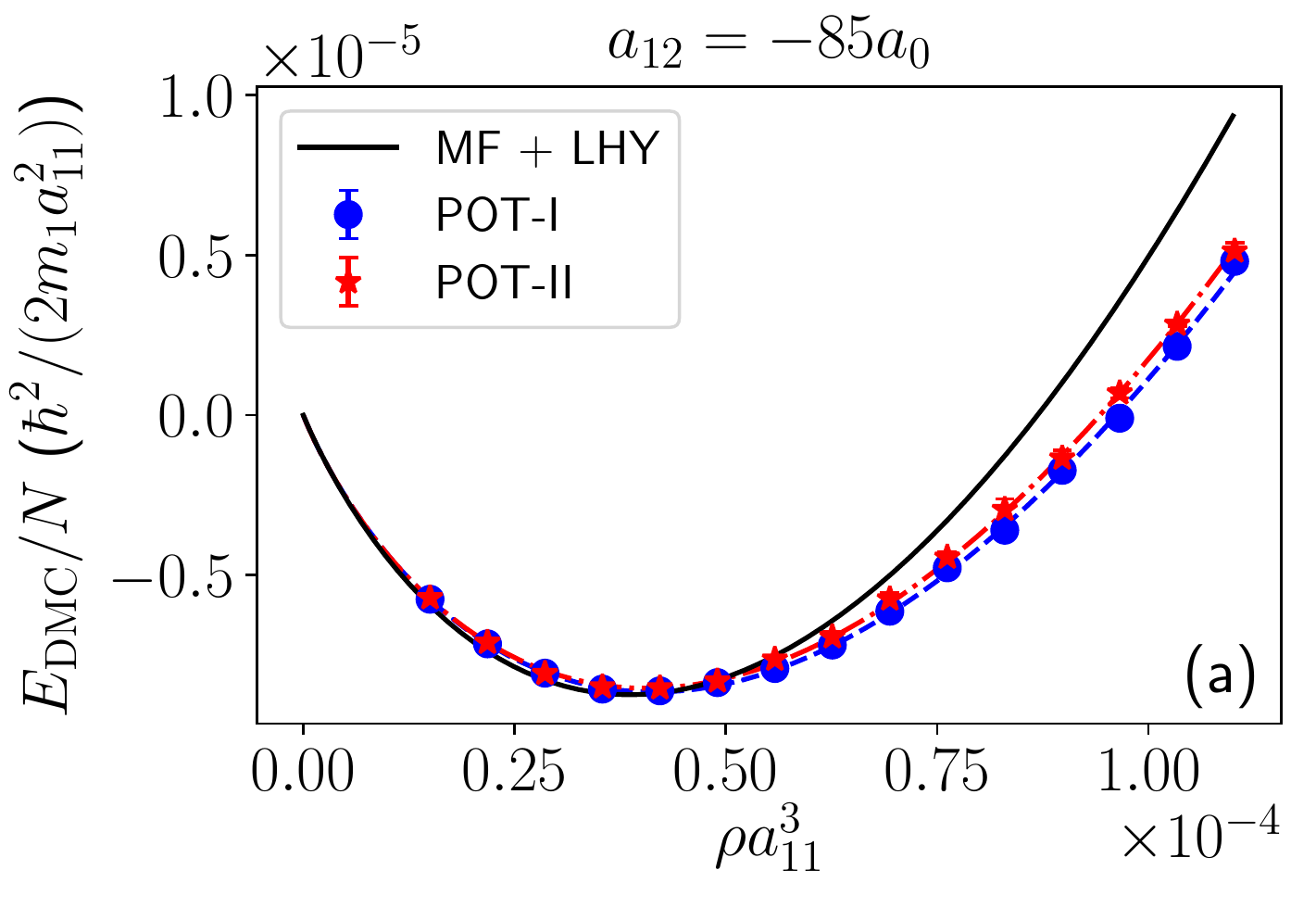}
		\includegraphics[width=\linewidth]{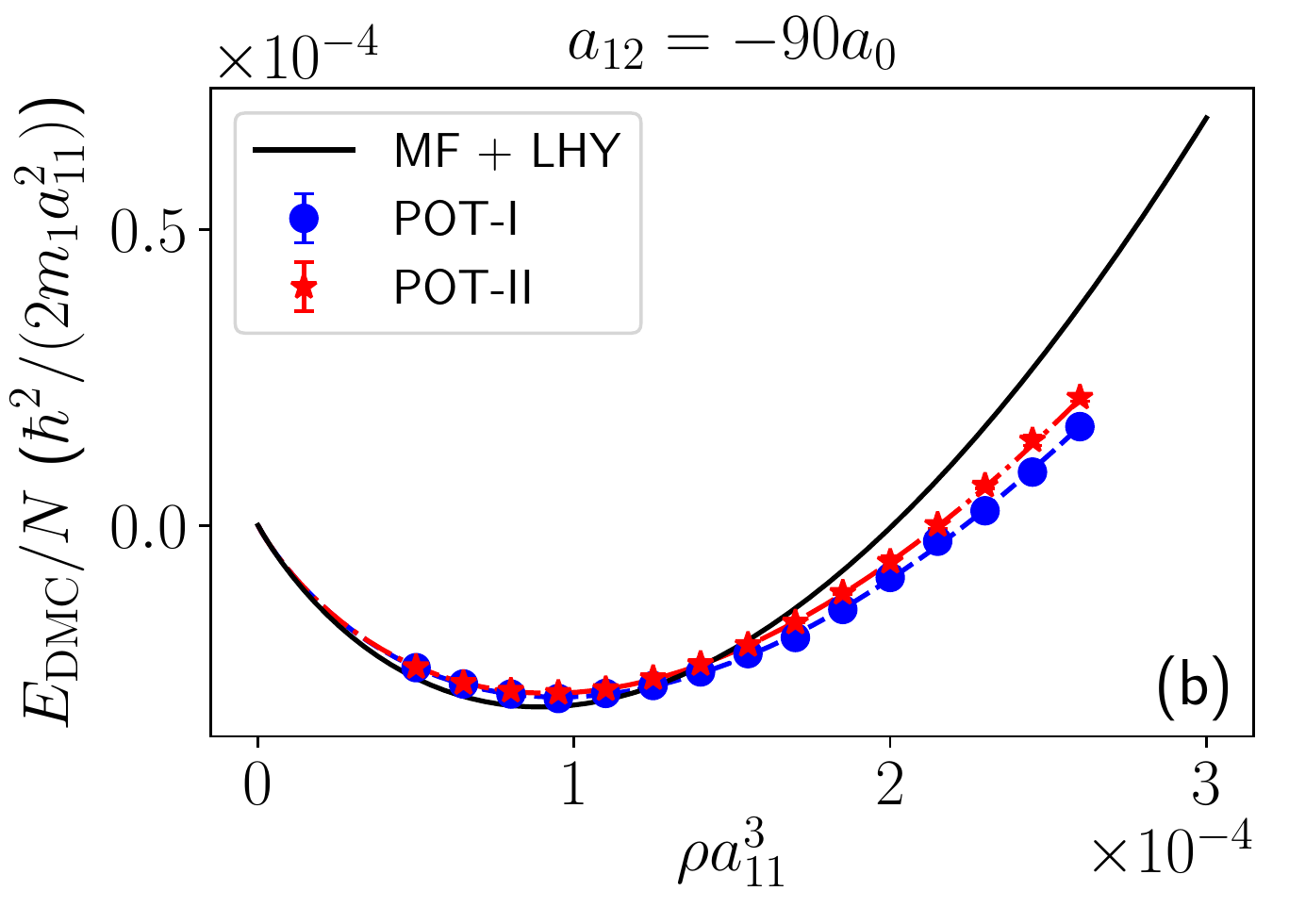}
		\includegraphics[width=\linewidth]{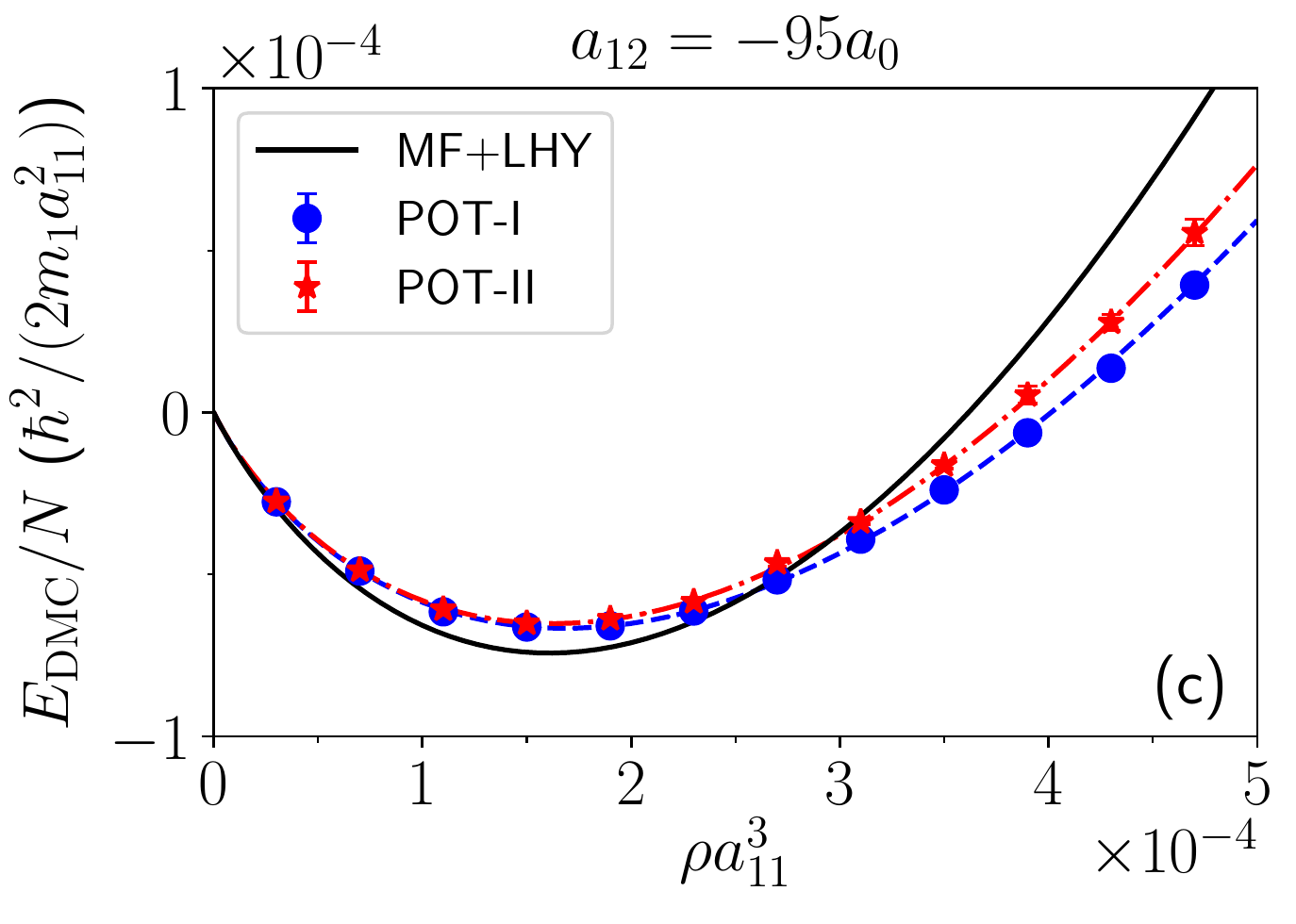}
		\caption{Energy per particle as a function of density, for values of $a_{12}=-85a_{0}$ (a), $a_{12}=-90a_{0}$ (b) and $a_{12}=-95a_{0}$ (c). DMC calculations are performed with POT-I and POT-II potentials, which satisfy both scattering parameters (see Eq. (\ref{eq:realistic-potentails}) and table (\ref{table:fit_params_realistic-pot}). Ratio of concentrations $\rho_2/\rho_1 = \sqrt{g_{11} / g_{22}}$ is kept fixed for each density, a criteria coming from the mean-field, also verified by QMC to predict the ground state.}
		\label{fig:universaleoscorrectareff}
	\end{figure}

	In Fig. (\ref{fig:universaleoscorrectareff}), we report the comparison 
between the DMC and MF+LHY equations of state, for three 
values of $a_{12}$ and using POT-I and POT-II as model potentials (see Eq. 
(\ref{eq:realistic-potentails}) and Table (\ref{table:realistic-pots_params})). 
For all three values of $a_{12}$, there is a slight increase in the equilibrium 
density, relative to the MF+LHY theory. This phenomenon was previously observed 
in a mixture with symmetric interactions 
\cite{cikojevic2019universality,petrov2015ultradilute,parisi2020quantum}, 
whereas in a $^{39}$K liquid mixture \cite{cikojevic2020finite}  
the significant increase of equilibrium density occurs because in that mixture the effective range is much 
larger than in the K-Rb one. In other words, in the K mixture the 
finite-range, beyond-LHY, negative energy contributions dominate. On the other 
hand, we observe that energies obtained with two different potentials but with 
the same $s$-wave scattering length and effective range collapses to a single 
equation of state, at least at densities not much larger than the equilibrium 
one. Therefore, the range of universality is extended, but now in terms of two 
scattering parameters \cite{cikojevic2020finite}. 
	
	\begin{table}[]
		\caption{$C_6$ coefficients of each channel and the corresponding scattering lengths and effective ranges.}
		\label{table:scattering_params}
		\centering
		\begin{tabular}{ c | c | c | c}
			Channel      & {$C_6$ (a. u.)} & $a / a_0$ & $r_{\rm eff} / a_0$                                                                       \\ \hline
			{K-K} (1-1) \cite{bohn1999collisional,derrico2007feshbach}   & {3897}  &  $65 $        & $168$                                                                                      \\ \hline
			{Rb-Rb} (2-2) \cite{marte2002feshbach} & {4707}  &  $ 100.4 $         & $153$          \\ \hline
			{K-Rb} (1-2) \cite{thalhammer2009collisional}  & {4285}         & $-85$ &   $795 $ \\
			&         & $-90$  &  $748 $ \\
			&         & $-95$ &   $707 $ \\
		\end{tabular}
	\end{table}

	\begin{table}[]
		\caption{ Parameters of POT-I and POT-II potentials which reproduce both scattering parameters (see Table {\ref{table:scattering_params}}). $R_0$ and $R_1$ are given in units $a_{11} = 65a_0$, and $V_0$ and $V_1$ are given in units $\hbar^2 / (m_1 a_{11}^2) = 10^{-3}{\rm Kelvin}$.}
		\label{table:realistic-pots_params}
		\centering
		\begin{tabular}{ c   c |  c  c  c  c}
			$a$ & Legend & $R_0$ & $R_1$ & $V_0$ & $V_1$ \\ \hline
			$ 65 a_0$         & POT-I     & $ 1.8797  $     & $0 $             & $ 1.2123 $     & $ 0 $ \\ 
			& POT-II     & $2.8386 $     & $ 3.5581 $     & $ 0.9140$     & $    0.3687$ \\ \hline
			$ 100.4 a_0$     & POT-I     & $ 2.7084 $     & $ 0 $         & $ 0.3337 $     & $ 0 $ \\ 
			& POT-II     & $ 3.8374 $     & $  4.8201 $     & $ 0.2366 $     & $ 0.0824 $ \\ \hline
			$-85 a_0$         & POT-I     & $  0 $         & $ 4.8897 $     & $ 0 $         & $ 0.01865$ \\ 
			& POT-II     & $  2.7513     $    & $ 3.6684 $     & $ 0.0632     $     & $  0.1189 $ \\ \hline
			$-90 a_0$         & POT-I     & $  0 $         & $ 4.8069 $     & $ 0 $         & $ 0.0205$ \\ 
			& POT-II     & $  2.7419     $     & $ 3.6558 $     & $ 0.0586     $     & $0.1197$ \\ \hline
			$-95 a_0$         & POT-I     & $ 0 $         & $ 4.7315 $     & $ 0 $         & $ 0.0222 $ \\ \hline
			& POT-II     & $ 2.7329 $     & $ 3.6438 $     & $ 0.0541 $     & $ 0.1205$ \\ 
		\end{tabular}
	\end{table}

	\begin{table}[tb]        
		\caption{Coefficients of fit $E/N = \alpha \rho a_{11}^3 + \beta (\rho a_{11}^3)^{\gamma}$ of the DMC energies per particle, obtained with  set of potentials POT-I and POT-II (se Eq. (\ref{eq:realistic-potentails} and Table (\ref{table:realistic-pots_params})). Comparison between different functionals are presented in Fig. (\ref{fig:universaleoscorrectareff}). Parameters $\alpha$ and $\beta$ are given in units $\hbar^2 / (m_1 a_{11}^2)=10^{-3}{\rm Kelvin}$, and $\gamma$ is adimensional. Scattering lengths in the repulsive channel are $a_{11} = 65 a_0$ and $a_{22} = 100.4 a_0$. For comparison, MF+LHY parameters are written in Table (\ref{table:fit_params_POT-SR}).}
		\label{table:fit_params_realistic-pot}
		\centering
		\resizebox{0.4\textwidth}{!}{    
			\begin{tabular}{ c  c c c c}
				$~a_{12}~$ &    Potentials & $\alpha$ & $\beta$ & $\gamma$\\ \hline                 
				\rule{0pt}{3ex}     -85     & POT-I  & -0.37  & 14.087 & 1.393 \\
				\rule{0pt}{3ex}             &    POT-II & -0.371 & 14.547 & 1.396 \\
				\rule{0pt}{3ex}     -90     & POT-I  & -0.578 & 11.778 & 1.359 \\
				\rule{0pt}{3ex}             &    POT-II & -0.573 & 12.894 & 1.369 \\
				\rule{0pt}{3ex}     -95     & POT-I  & -0.824 & 9.752  & 1.316 \\
				\rule{0pt}{3ex}             &    POT-II & -0.774 & 12.22  & 1.351 \\
			\end{tabular}
		}
	\end{table}

\section{\label{sec:dft_results}Density-functional results}

The stability of the self-bound mixture of $^{41}$K-$^{87}$Rb in 
free space implies the presence of a surface and a positive surface 
tension associated to it. We studied the surface properties of spherical $^{41}$K-$^{87}$Rb
droplets within the DFT approach
in the MF+LHY framework, 
for a fixed ratio of densities $\rho_2/\rho_1=\sqrt{g_{11}/g_{22}} $
corresponding to the equilibrium one for the homogeneous mixture.
We provide additional results using a more accurate density functional
obtained from the ab-initio QMC results discussed in the
previous Section.

We have first studied a planar surface, with the aim of determining the 
surface tension $\sigma $ in a range of values of the inter-particle 
attractive interaction $a_{12}$ accessible to the experiments.
When studying a planar surface, it is useful to use a slab geometry, i.e.,
we assume an extended homogeneous system in the $xy$ plane 
(with periodic boundary conditions), 
and with a finite extension along the $z$-direction. In this direction,
two liquid-vacuum interfaces are formed, with the slab width 
thick enough to have a constant density region between the
two confining surfaces ("bulk" phase).
This amounts to neglect curvature effects, which will be
explicitly considered later on.
The surface width of the density profile along $z$ 
can be quantified by the parameter $\Delta$, which measures 
the width between surface points at $90\%$ and 
 $10\%$ of the bulk total density.

By defining the following coefficients,
\begin{equation}\label{alpha}
        C_{\rm K}  = \frac{1}{4}\left( 
\frac{\hbar^2}{2m_1}+\frac{\hbar^2}{2m_2}\sqrt{\frac{g_{11}}{g_{22}}}\right) \ ,
        \end{equation}
        \begin{equation}\label{beta}
        C_\delta = g_{11}+ g_{12}\sqrt{\frac{g_{11}}{g_{22}}} \ ,
        \end{equation}
        \begin{equation}\label{gamma}
        C_\rho = \frac{8}{15 
\pi^2}\left(\frac{m_1}{\hbar^2}\right)^{3/2}g_{11}^{5/2} 
\left[1+\left(\frac{m_1}{m_2}\right)^{3/5}\sqrt{\frac{g_{22}}{g_{11}}}\right] \ 
,
\end{equation}
the effective single-component energy density of the mixture within the MF+LHY theory,
expressed for simplicity in terms of the density $\rho_1$, reads
\begin{equation} \label{en_dens}
\mathcal{E} = C_{\rm K}\frac{(\nabla \rho_1)^2}{\rho_1} + C_\delta \rho_1^2 + 
C_\rho \rho_1^{5/2}.
\end{equation}
We recall that knowledge of one density is enough to characterize
the whole droplet because of the underlying assumption
$\rho_2/\rho_1 = \sqrt{g_{11}/g_{22}}$~\cite{Ancilotto_2018self,staudinger2018self}.

Remarkably, the surface tension of the planar 
interface described by local energy functionals of the form (\ref{en_dens})
can be estimated, without any prior knowledge of the density profile, 
by calculating the following integral \cite{stringari_treiner},
\begin{equation}\label{sigma}
    \sigma =2 \int_{0}^{\rho_0} d\rho_1\, \sqrt{C_{\rm 
K}\left(C_{\delta}\rho_1+C_\rho \rho_1^{3/2}-\mu_0\right)}\, ,
\end{equation}
where $\mu_0=C_\delta \rho_1 +C_\rho \rho_1^{3/2}$ is the chemical potential of 
a liquid system 
in equilibrium with the vacuum, evaluated at the equilibrium density $\rho _0$.
The density profile can also be obtained by simple quadrature, 
solving the implicit equation 
\begin{equation}\label{profile}
  z(\rho )=z_0+\int _{\rho_0/2}^{\rho} \frac {1}{h(\rho ^\prime)}d\rho ^\prime
\end{equation}
where $\rho (z_0)=\rho_0/2$ and
\begin{equation}
    h(\rho) = - \sqrt{\left(\frac{\rho}{C_{\rm 
K}}\right)\left[C_\delta\rho^2+C_\rho \rho^{5/2}-\mu_0\rho\right]} 
\end{equation}
Here $\rho $ can have any value in the interval $[0,\rho_0]$.

\begin{figure}
    \centering
    \includegraphics[width=\linewidth]{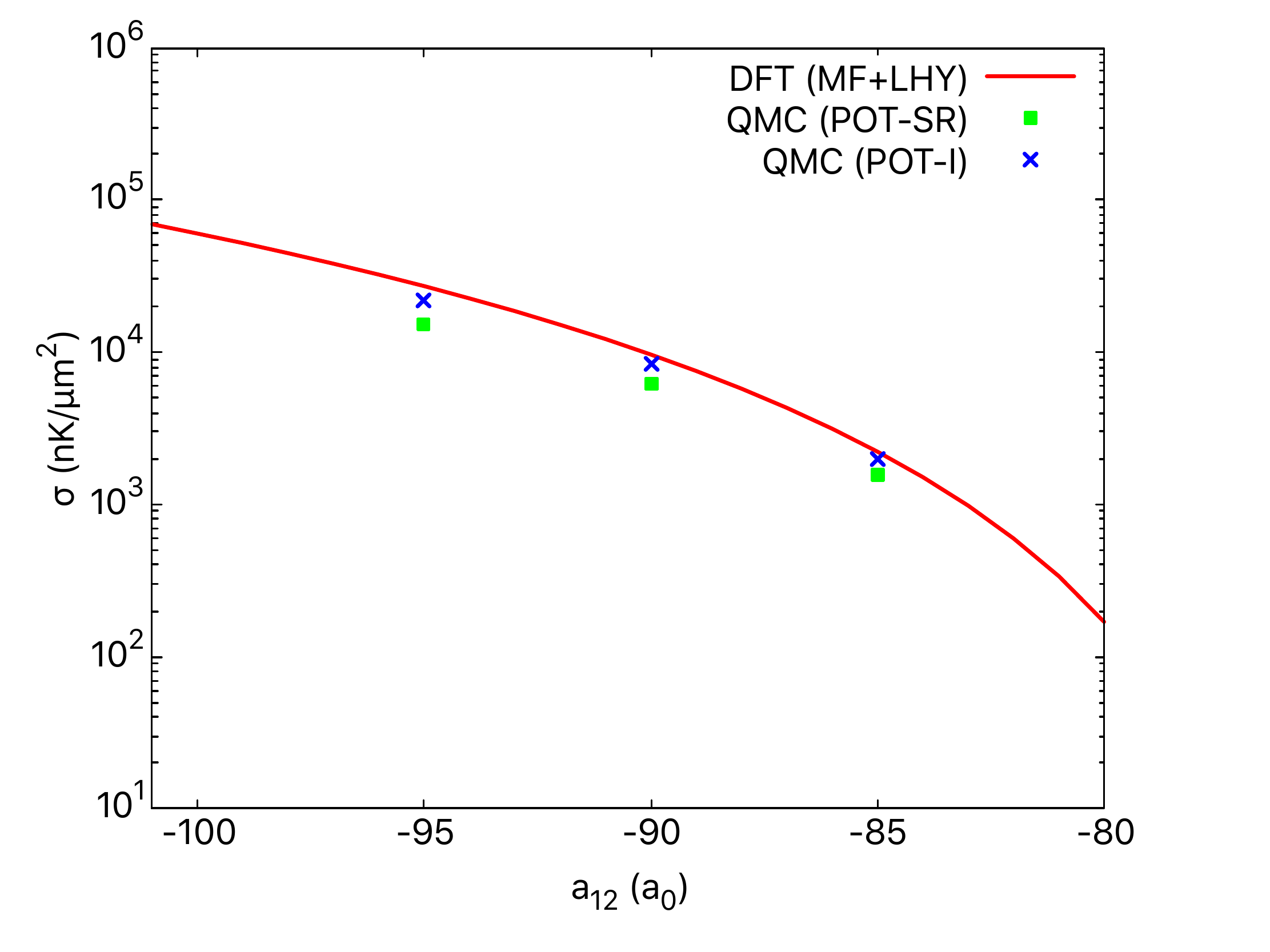}
    \caption{Surface tension as a function of the (attractive) inter-species scattering length $a_{12}$. The solid red line is obtained using the MF+LHY functional, green squares and blue crosses are the predictions assuming POT-SR and POT-I functionals, respectively (see Table \ref{table:fit_params_POT-SR} and \ref{table:fit_params_realistic-pot}).}
    \label{fig:surftens}
\end{figure}

The calculated surface tension for different values of $a_{12}$ is shown 
in Fig.~\ref{fig:surftens}, and compared with the results obtained from the
QMC-based functional. Notice that relatively small changes in the 
inter-species interaction strength cause order-of-magnitude changes
in the surface tension, which is highlighted by the logarithmic scale introduced in the $\sigma$-axis. Surface tension obtained with the QMC-based functionals POT-SR and POT-I are both below the predictions of the MF+LHY functional, with POT-SR having larger deviations from MF+LHY.

The width of the liquid-vacuum interface profile is strongly dependent 
on the inter-species scattering length $a_{12}$. In a droplet this will
affect the overall shape of the droplet itself, depending upon the
value of the total number of particles $N$.
For larger values of $N$ the droplet will be characterized
by a central region of fairly uniform  
density (``bulk") and an external surface region 
where the density drops to zero 
with the distance from its center, whereas it will be 
an ``all surface", gaussian-like droplet, where 
the central bulk region is almost absent, for low values of $N$.
By defining the droplet radius $R$ as
\begin{equation}\label{defR}
     R = \sqrt[3]{\frac{3N}{4\pi\rho_{\text{bulk}}}},\,
\end{equation}
in the first case the 
ratio between the surface width $\Delta$ and the droplet radius $R$ 
is $\Delta/R \ll 1$, while in the second case 
$\Delta/R \gg 1$.

We must notice that not all values of $N$ are 
allowed in a droplet for a given $a_{12}$ because small droplets, i.e., those with a 
number of particle below some critical value $N_c$ become
unstable when the kinetic energy dominates over the 
interaction energy, eventually causing the evaporation of the droplet.
In order to estimate the critical size $N^c$ we 
make a simple variational ansatz for the 
radial density profile, which is 
a good approximation for spherical, small droplets:
\begin{equation}
\label{variat}
    \rho_1(r) = \frac {N_1}{\pi ^{3/2}\sigma ^3}e^{-r^2/\sigma ^2}.
\end{equation}

\begin{figure}
\centering
\includegraphics[width=\linewidth]{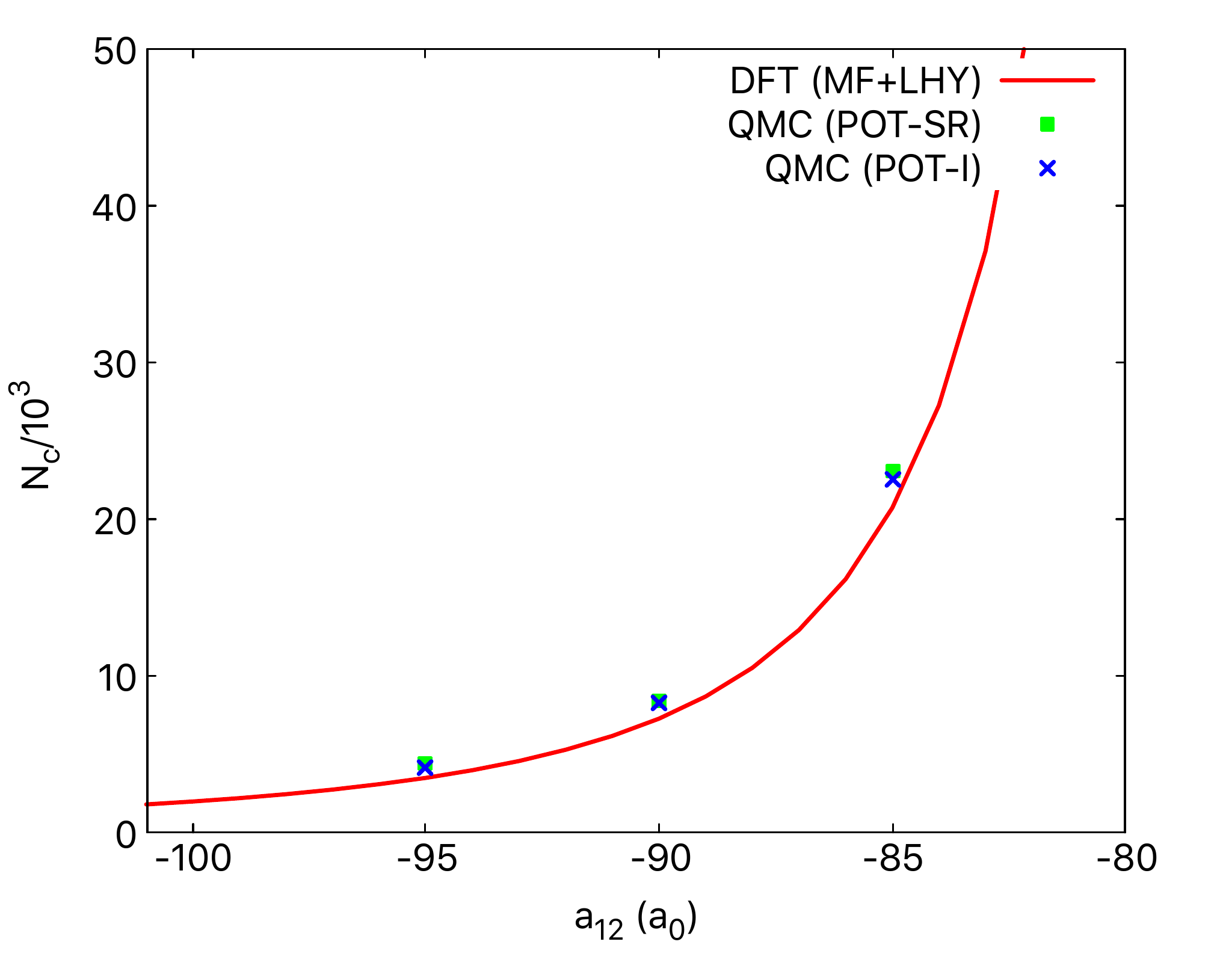}
\caption{Critical atom number $N_c$ as a function of the inter-species scattering length. The solid red line is obtained using the MF+LHY functional, green squares and blue crosses are the predictions assuming POT-SR and POT-I functionals, respectively (see Table \ref{table:fit_params_POT-SR} and \ref{table:fit_params_realistic-pot}).}
\label{fig:critical_atom_number}
\end{figure}

We use the above ansatz in the energy functional \eqref{en_dens} and 
impose the condition for a minimum,
$\partial E /\partial \sigma =0$, together with the additional 
requirement $E =0$ which marks the line separating
stable droplets with negative total energies from 
unstable ones with positive energies.
Solving for the total number of atoms of species 1, $N_1$, we 
find the critical droplet size $N_c=N_{c1}(1+\sqrt{g_{11}/g_{22}})$,
where
\begin{equation}
\label{ncrit}
    N_{c1} = -\frac {c}{\lambda ^{9/2}(a/\lambda^2+b/\lambda^3)} 
\end{equation}
Here $\lambda=-3b/5a$, with $a,b,c$ being given by the expressions
$a=6 C_{\rm K}$, $b=C_\delta /(2\pi^{3/2})$ and $c=4 C_\rho 
/(5\sqrt{10}\pi^{9/4})$.
The table \ref{table:summary} and the figure 
\ref{fig:critical_atom_number} show the critical atom number 
for some values of the scattering length computed with DFT and 
QMC methods.

\begin{figure}[htbp!]
\centering
\includegraphics[width=\linewidth]{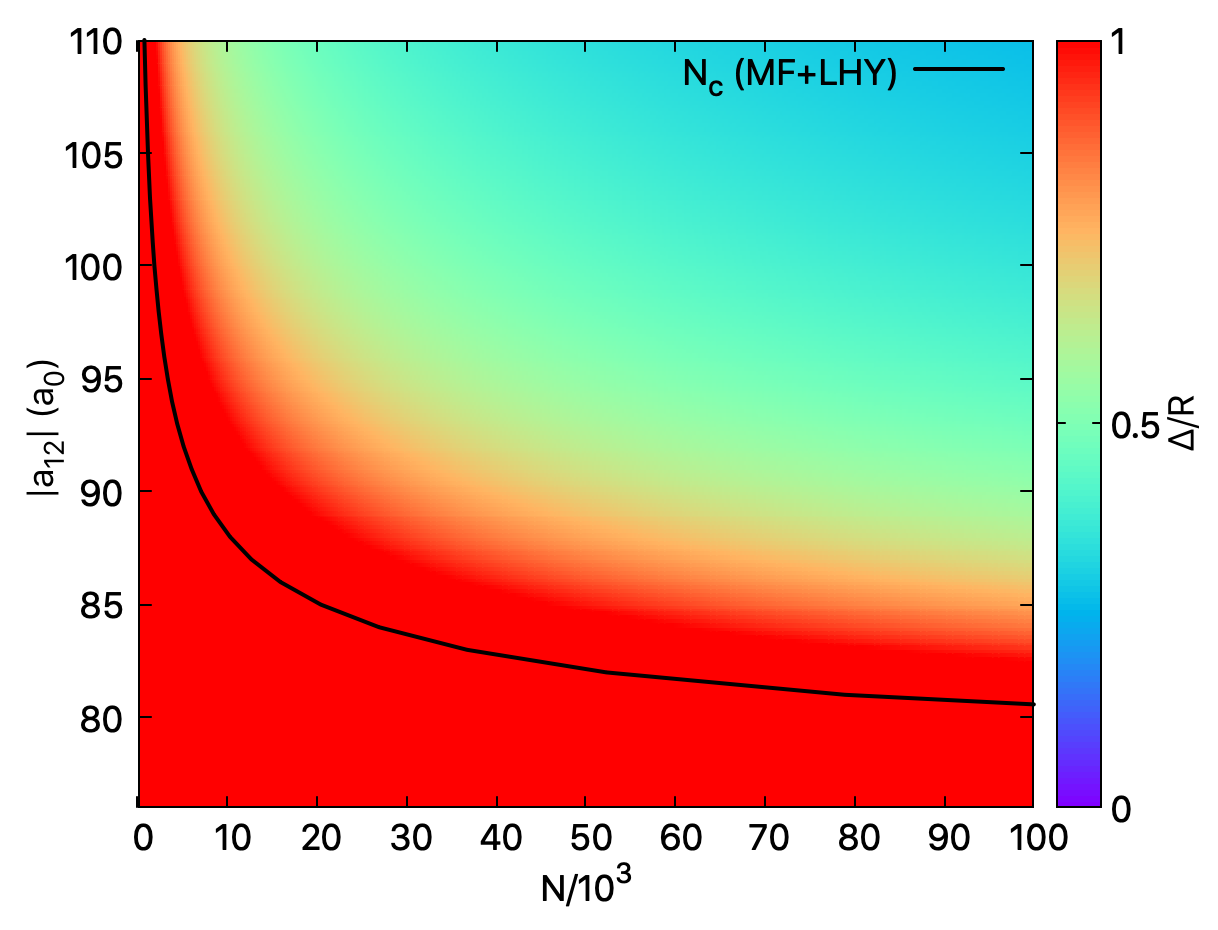}
\caption{ 
The ratio $\Delta/R$ is represented as a function of $|a_{12}|$ and $N$. 
The red region identifies gaussian-like radial density profiles while 
the green and blue region identifies droplets with a central bulk region. 
The black solid line marks the total critical number of particles 
$N_c$ below which the droplet evaporates.}
\label{fig:colormap}
\end{figure}

Figure \ref{fig:colormap} summarizes our results: 
the color scales shows the ratio between the surface width $\Delta$ 
and the radius $R$ of the droplet, which evaporates for a number of 
particles below the critical value (solid black line). The red region 
identifies gaussian-like, "all-surface" droplets, while the green and 
blue region identifies droplets with a well defined, central "bulk" density. 
Notice that the black line belongs to the red region for each $|a_{12}|$ 
so the assumption of a gaussian density profile for the variational 
study of the critical atom number is indeed justified. 

The surface tension for the $^{41}$K - $^{87}$Rb self-bound mixture
has been obtained for a planar interface (we will call it $\sigma_0$ to 
distinguish it from the size-dependent surface tension of a droplet), 
though in real droplets an interfacial curvature of the 
surface in contact with the vacuum is present. 
The curvature-dependent surface tension can be expressed in term
of the so-called Tolman length $\delta$~\cite{tolman1949}.
To a first approximation 
the Tolman length $\delta$ 
is independent of the droplet size, and it gives the
size-dependent surface tension in term of the one for a planar surface
\cite{thermo_delta, tolmann}
\begin{equation}\label{sigmacorr}
\sigma(R) =\sigma_0\left(1-\frac{2\delta}{R}\right).
\end{equation}
A thermodynamic argument relates the Tolman length to
the isothermal compressibility $\kappa^{-1}$ and surface tension $\sigma_0$ \cite{bartell2001tolman}
\begin{equation}
\label{tolman}
\delta \approx -\kappa^{-1}\sigma_0 .
\end{equation} 

The isothermal compressibility $\kappa^{-1}$ of a self-bound quantum 
mixture with equilibrium bulk densities $\rho_1$ and $\rho_2$ in the 
MF+LHY approach is given by \cite{Ancilotto_2018self}
\begin{equation}
\kappa^{-1} = \left( g_{11}\rho_{1}^2+ g_{22}\rho_{2}^2+2g_{12}\rho_1\rho_2 + 
\frac{15}{4}\mathcal{E}_{\text{LHY}} \right)^{-1}.
\end{equation}
Previously, some studies have noticed that the product $\kappa^{-1} \sigma_0$ 
is a fundamental characteristic length in liquid 
droplets \cite{ksigma1,ksigma2,ksigma3,ksigma4,ksigma5} though 
the connection with the Tolman length $\delta$ was never explicitly made. 
We checked the validity of Eq. (\ref{tolman}) by independently 
computing $\delta$ using the Liquid Drop Model (LDM), 
i.e., writing the calculated total energy of a droplet 
made of $N$ atoms as
\begin{equation}
\label{ldm}
E=aN+bN^{2/3}+cN^{1/3}
\end{equation}
where the separate bulk, surface and curvature
contributions to the total energy of the droplet are
highlighted.
By using the relation (\ref{sigmacorr})
one can see that $c=8\pi(3/4\pi \rho_0)^{1/3}\sigma_0 \delta$.
The coefficient $c$ is in turn obtained by fitting the
calculated energies using the LDM expression quoted above,
allowing
to determine the Tolman length $\delta$. Predictions for the Tolman length are summarized in Table \ref{table:summary}.

\begin{figure}
\centering
\includegraphics[width=\linewidth]{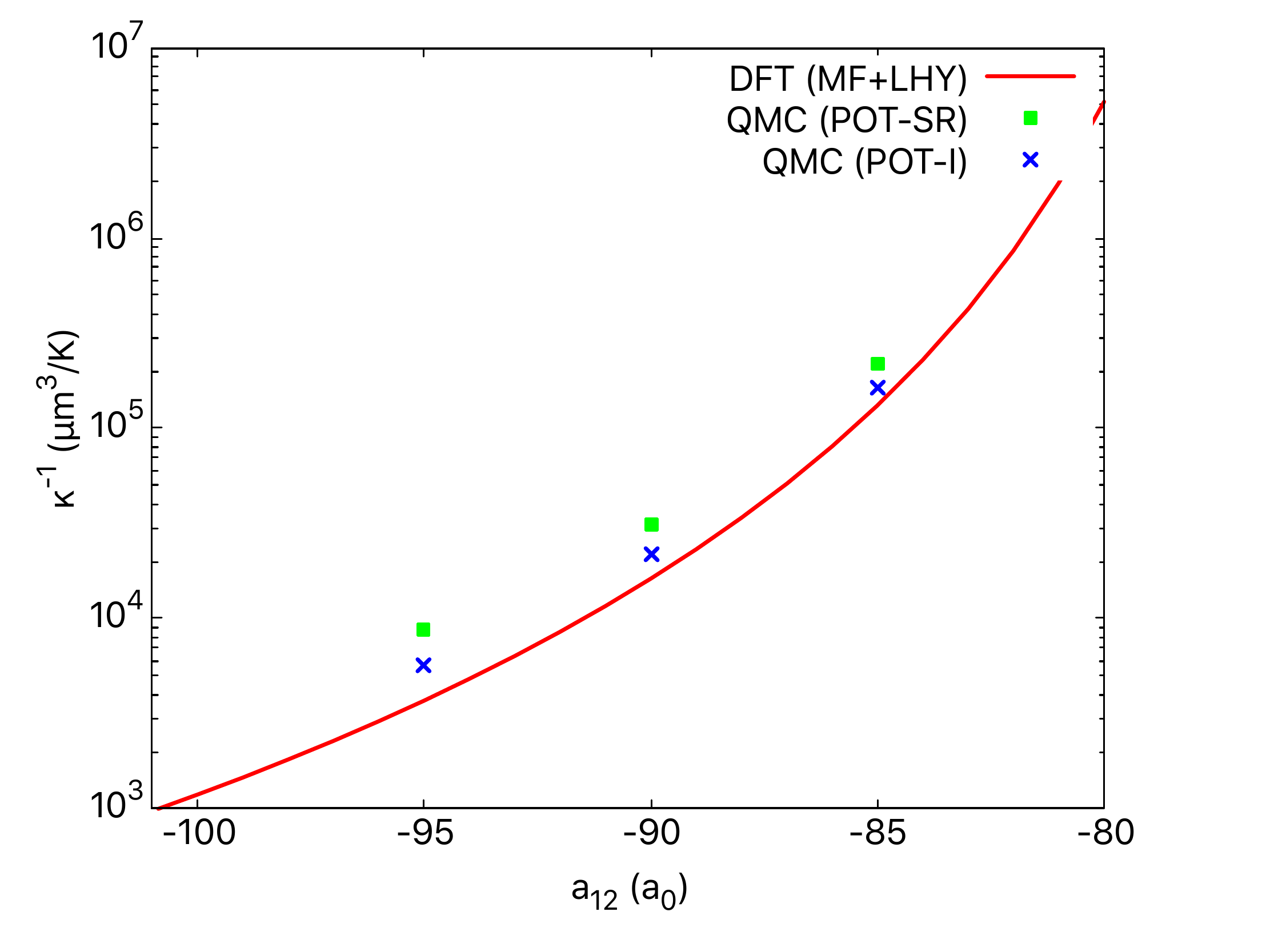}
\caption{Compressibility $\kappa^{-1}$ as a function of the inter-species 
scattering length $a_{12}$ in the self-bound droplet regime. The solid red line is obtained using the MF+LHY functional, green squares and blue crosses are the predictions assuming POT-SR and POT-I functionals, respectively (see Table \ref{table:fit_params_POT-SR} and \ref{table:fit_params_realistic-pot}).}
\label{fig:compressibility}
\end{figure}

In Fig.~\ref{fig:compressibility},  the compressibility computed with the 
MF+LHY approach is shown for a set 
of values of the scattering length and compared to QMC results. In contrast to $^{39}$K droplets \cite{cikojevic2020qmcbased}, the compressibility of QMC-based functional is higher compared to the MF+LHY, due to relatively smaller values of effective ranges in a K-Rb mixture.

 \begin{table*}[]
 	\caption{Summary of all the quantities reported in the paper. Mass $m$ is the mass of $^{41}$Rb atom, and $a_0$ is Bohr radius. $\epsilon_r$(POT-SR) and $\epsilon_r$(POT-I) are the relative differences $ (O^{\rm POT-SR} - O^{\rm MF+LHY}) / |O^{\rm MF+LHY}|$ and $ (O^{\rm POT-I} - O^{\rm MF+LHY}) / |O^{\rm MF+LHY}|$ for observable $O$, given in percentages. Positive (negative) values of $\epsilon_r$ mean that QMC functionals predict higher (lower) values of the observable with respect to MF+LHY one.}
 	\label{table:summary}
 	\centering
 	\begin{tabular}{ c | c |  c | c | c | c | c}
 		Observable & $a_{12}/ a_0$ & MF+LHY &   POT-SR & POT-I &  $\epsilon_r$(POT-SR)   &  $\epsilon_r$(POT-I) \Tstrut \\ [3pt] \hline 				
 		$  $ & $ -85 $ & $ 2.07 \cdot 10^4 $ & $ 2.31 \cdot 10^4 $ & $ 2.26 \cdot 10^4 $ & $ 12 $ & $ 9$\Tstrut \\ [3pt]
 		$N_c$ & $ -90 $ & $ 7.28 \cdot 10^3 $ & $ 8.47 \cdot 10^3 $ & $ 8.31 \cdot 10^3 $ & $ 16 $ & $ 14$\\ [3pt]
 		$  $ & $ -95 $ & $ 3.49 \cdot 10^3 $ & $ 4.44 \cdot 10^3 $ & $ 4.10 \cdot 10^3 $ & $ 27 $ & $ 17$\\ [3pt]
 		\hline$  $ & $ -85 $ & $ 1.97 \cdot 10^{-20} $ & $ 1.40 \cdot 10^{-20} $ & $ 1.77 \cdot 10^{-20} $ & $ -29 $ & $ -10$\Tstrut \\ [3pt]
 		$\dfrac{\sigma}{\hbar^2 / (m a_0^4)}$ & $ -90 $ & $ 8.55 \cdot 10^{-20} $ & $ 5.45 \cdot 10^{-20} $ & $ 7.42 \cdot 10^{-20} $ & $ -36 $ & $ -13$\\ [3pt]
 		$  $ & $ -95 $ & $ 2.40 \cdot 10^{-19} $ & $ 1.34 \cdot 10^{-19} $ & $ 1.95 \cdot 10^{-19} $ & $ -44 $ & $ -19$\\ [3pt]
 		\hline$  $ & $ -85 $ & $ 4.17 \cdot 10^4 $ & $ 4.78 \cdot 10^4 $ & $ 4.48 \cdot 10^4 $ & $ 14 $ & $ 7$\Tstrut\\ [3pt]
 		$\Delta/a_0$ & $ -90 $ & $ 2.23 \cdot 10^4 $ & $ 2.65 \cdot 10^4 $ & $ 2.46 \cdot 10^4 $ & $ 19 $ & $ 11$\\[3pt]
 		$  $ & $ -95 $ & $ 1.43 \cdot 10^4 $ & $ 1.80 \cdot 10^4 $ & $ 1.65 \cdot 10^4 $ & $ 26 $ & $ 15$\\[3pt]
 		\hline$   $ & $ -85 $ & $ -5.53 \cdot 10^3 $ & $ -6.57 \cdot 10^3 $ & $ -6.16 \cdot 10^3 $ & $ -19 $ & $ -11$\Tstrut\\[3pt]
 		${\delta}/{a_0} $ & $ -90 $ & $ -2.95 \cdot 10^3 $ & $ -3.67 \cdot 10^3 $ & $ -3.43 \cdot 10^3 $ & $ -24 $ & $ -16$\\[3pt]
 		$   $ & $ -95 $ & $ -1.90 \cdot 10^3 $ & $ -2.50 \cdot 10^3 $ & $ -2.35 \cdot 10^3 $ & $ -32 $ & $ -24$\\[3pt]
 		\hline $    $ & $ -85 $ & $ 2.81 \cdot 10^{23} $ & $ 4.69 \cdot 10^{23} $ & $ 3.47 \cdot 10^{23} $ & $ 67 $ & $ 24$\Tstrut\\
 		$\dfrac{\kappa^{-1}}{m a_0^5 / \hbar^2}  $ & $ -90 $ & $ 3.45 \cdot 10^{22} $ & $ 6.73 \cdot 10^{22} $ & $ 4.62 \cdot 10^{22} $ & $ 95 $ & $ 34$\\
 		$    $ & $ -95 $ & $ 7.90 \cdot 10^{21} $ & $ 1.86 \cdot 10^{22} $ & $ 1.20 \cdot 10^{22} $ & $ 135 $ & $ 52$\\[3pt]		
 	\end{tabular}
 \end{table*}

\begin{figure}
\centering
\includegraphics[width=\linewidth]{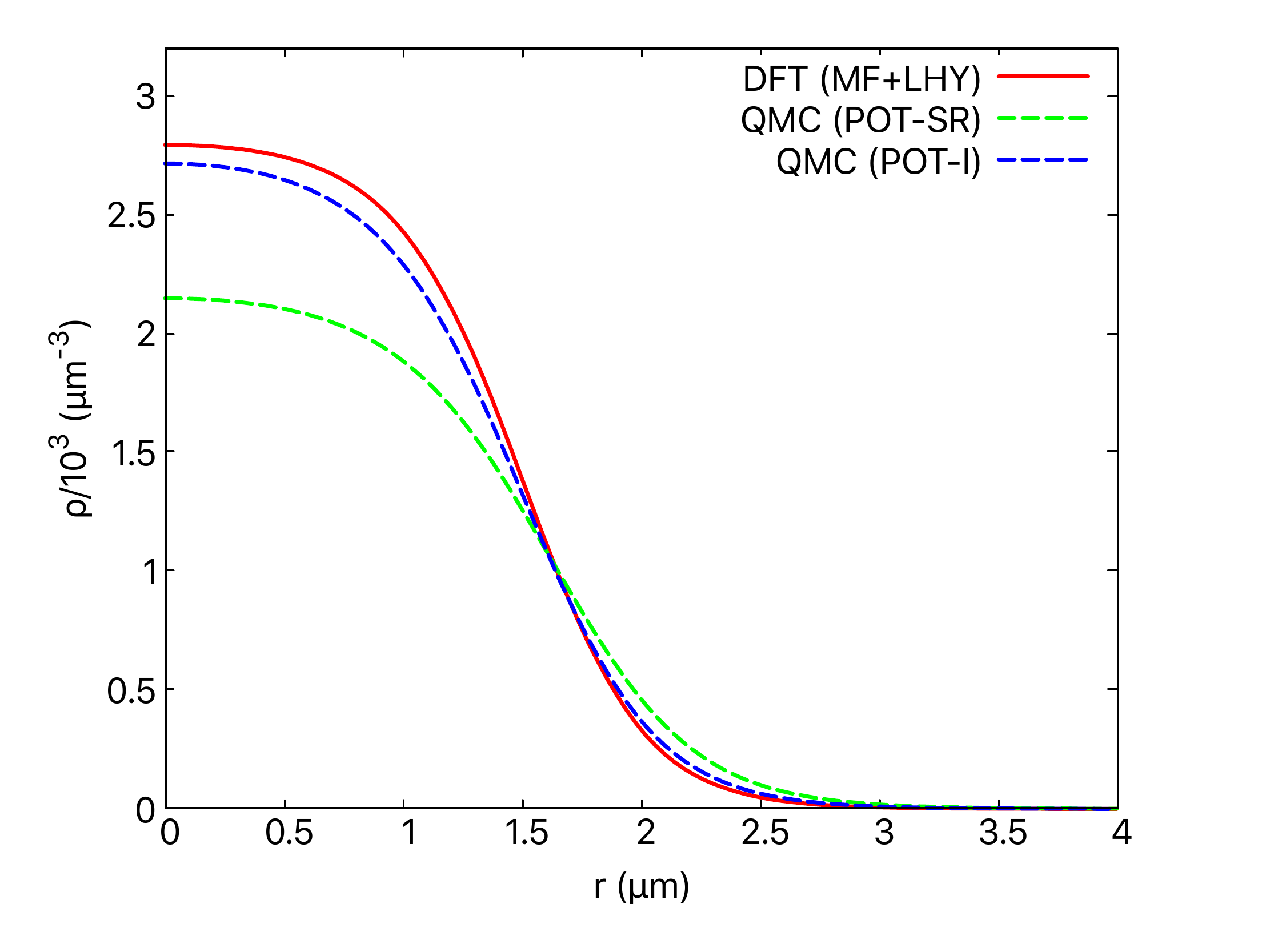}
\caption{Radial density profile for a droplet with $a_{12}=-90\,a_0$:
the prediction of MF+LHY (red solid line) are compared with the results obtained
from a density functional reproducing the QMC results, either at
zero-range (POT-SR, green dashed line) and with finite range effects (POT-I, blue dashed line). 
}
\label{fig:droplet}
\end{figure}

\begin{figure}
\centering
\includegraphics[width=\linewidth]{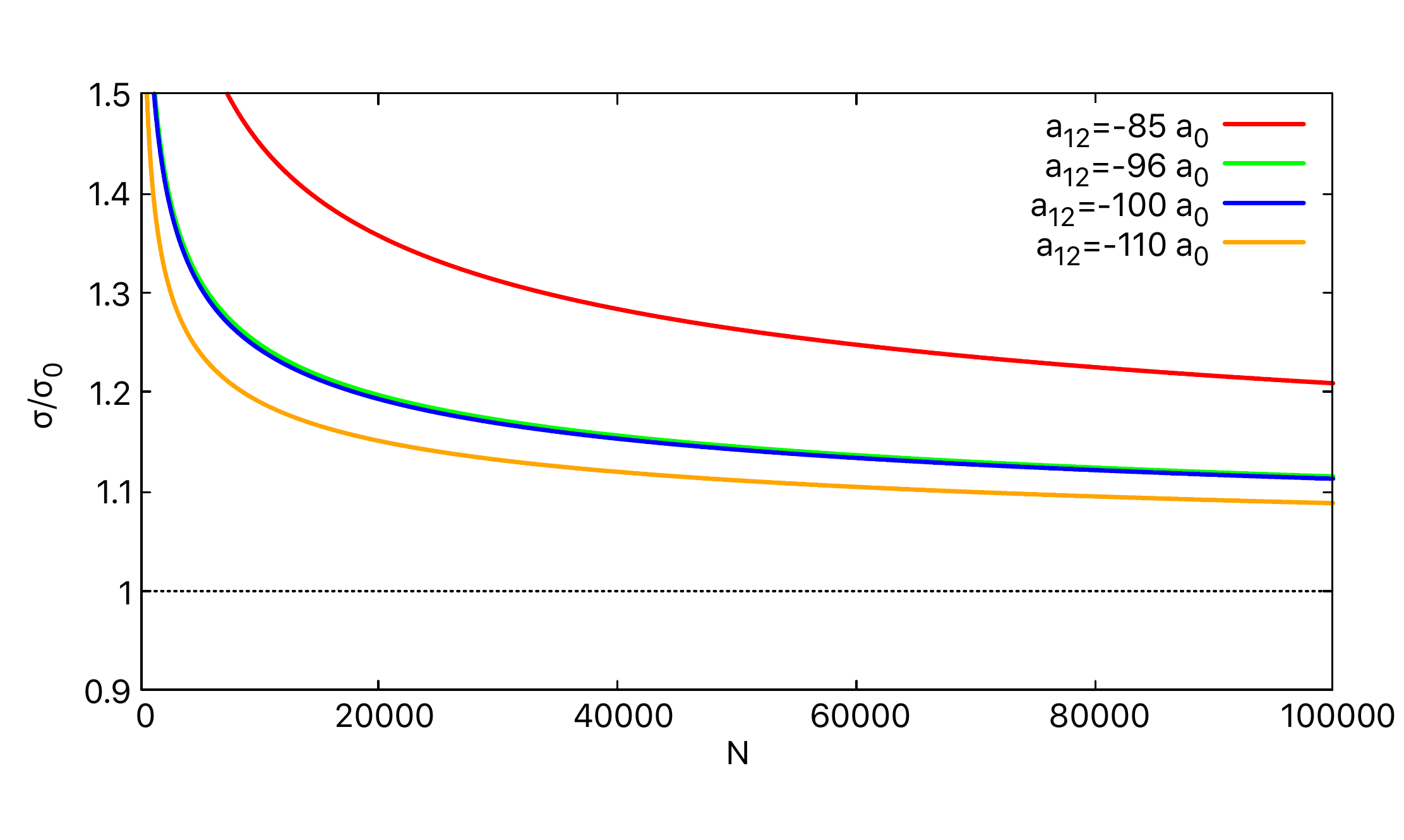}
\caption{Corrected and normalized surface tension $\sigma/\sigma_0$ as a function of the total number of particles $N$ in the droplet for a set of values of the inter-species scattering length $a_{12}$, computed within the MF+LHY framework. }
\label{fig:sizecorr_surf}
\end{figure}

Fig. \ref{fig:droplet} shows the calculated radial density profile
for a droplet with $a_{12}=-90\,a_0$, where a comparison is
made between the prediction of the MF+LHY approach and those from the 
QMC-based energy functionals. The QMC-based functional, which relies on  
short-range model potentials (POT-SR, see Fig. 
\ref{fig:universaleos2onlyshortrange}), predicts less dense droplets due 
to repulsive beyond-LHY energy contributions. Profiles obtained with the POT-I and MF+LHY functionals are more similar since the effect of increasing the effective range, included in POT-I through model potentials, is to increase binding energy and the peak density.

In Fig.~\ref{fig:sizecorr_surf}, we show the dependence of the corrected 
surface tension $\sigma$ as a function of the droplet’s total number of 
particles $N$, computed within the MF+LHY framework by combining Eq. \eqref{sigmacorr} with Eq. \eqref{defR}.

In Table \ref{table:summary}, we summarize the predictions for all observables 
analyzed 
in our work with the MF+LHY and QMC-based functionals for $a_{12}=-85a_0$, 
$-90a_0$ and $-95a_0$. It can be seen that QMC functionals show increasing 
deviation from the predictions of the MF+LHY functional as $|a_{12}|$ increases. 
QMC-based density functionals constructed assuming short-range model potentials 
(POT-SR) predict larger deviations from the MF+LHY theory. When the correct 
effective range is included in the model potentials (POT-I), the predictions 
become more similar to those of MF+LHY. Variable most sensitive  to beyond-LHY 
energy corrections appears to be the compressibility $\kappa^{-1}$, with a 
relative difference in the range $\approx 20\%$ to $130\%$. This could have an 
impact on the collective excitation modes of a droplet \cite{hu2020collective}, 
which will be investigated in a future study.

\section{\label{sec:summary}Conclusion}

We have performed QMC calculations for the ground state of a K-Rb liquid 
mixture. We found that the zero-range model potentials used in a QMC calculation 
predict significant beyond LHY energy contributions. Otherwise, when the 
$s$-wave effective range is included in the model potentials, the energies fall 
close to the MF+LHY energies. Using both the MF+LHY and the QMC-based 
functionals, we have investigated fundamental surface properties for an 
experimentally relevant range of scattering parameters. These properties are 
relevant to the ongoing experiments because the observed droplets have a large 
surface-to-volume ratio. Upon entering a more correlated (denser) regime, the 
differences between the predictions for all quantities with the MF+LHY and the 
QMC-based functional grow. \\

	\acknowledgments

	This work has been supported by the Ministerio de Economia, Industria y Competitividad (MINECO, Spain) under grants
	Nos. FIS2017-84114-C2-1-P and FIS2017-87801-P (AEI/FEDER, UE), and by the EC Research Innovation Action under the H2020 Programme,
	Project HPC-EUROPA3 (INFRAIA-2016-1-730897). V. C. gratefully acknowledges the
	computer resources and technical support provided by Barcelona Supercomputing Center.
	We also acknowledge financial support from Secretaria d'Universitats i Recerca del Departament d'Empresa i Coneixement de la Generalitat de Catalunya, co-funded by the European Union Regional Development Fund within the ERDF Operational Program of Catalunya (project QuantumCat, ref. 001-P-001644).

\bibliography{references.bib} 
	
\end{document}